\documentclass[letter,english]{aa}
\usepackage{times}
\usepackage[T1]{fontenc}
\usepackage[latin1]{inputenc}
\usepackage{graphicx}
\usepackage{amssymb}
\usepackage{amsmath}
\usepackage[authoryear]{natbib}
\makeatletter

\usepackage{babel}
\makeatother
\begin{document}

\title{Might Carbon-Atmosphere White Dwarfs Harbour\\
 a New Type of Pulsating Star?
}

\author{G. Fontaine\inst{1}
\and P. Brassard\inst{1}
\and P. Dufour\inst{2}
}

\institute{
D\'epartement de Physique, Universit\'e de Montr\'eal, C.P. 6128,
  Succursale Centre-Ville, Montr\'eal, QC H3C 3J7, Canada
\\ \email{fontaine,brassard@astro.umontreal.ca}\\
\and Department of Astronomy and Steward Observatory, University of
Arizona, 933 North Cherry Avenue, Tucson, AZ 85721, USA
\\ \email{dufourpa@as.arizona.edu}
}

\offprints{G. Fontaine}

\date{Received 26 February 2008 / Accepted 14 March 2008}

\newpage

\abstract{}
{In the light of the recent and unexpected discovery of a brand new type
of white dwarfs, those with carbon-dominated atmospheres, we examine the
asteroseismological potential of such stars. The motivation behind this
is based on the observation that past models of carbon-atmosphere white
dwarfs have partially ionized outer layers that bear strong resemblance
with those responsible for mode excitation in models of pulsating DB
(helium-atmosphere) and pulsating DA (hydrogen-atmosphere) white dwarfs.}
{We carry out a stability survey of models of carbon-atmosphere white
dwarfs following a full nonadiabatic approach. To connect with
previously known pulsating white dwarfs, we specifically search for
excited $g$-modes with $\ell = 1$ in the period window 80-1200 s. We
concentrate on models with $T_{\rm eff}$ $\leq$ 30,000 K, a limit below
which the real carbon-atmosphere stars are found, and investigate the
effects of changing the surface gravity, the composition of the
envelope, and the treatment of the convective efficiency.}
{Our exciting main result is that carbon-atmosphere white dwarfs may
indeed drive low-order $g$-modes in certain regions of parameter
space. For instance, log $g$ = 8.0 models characterized by an envelope
composition $X$(C) = $X$(He) = 0.5 and ML2 convection, show a broad
instability strip in the range 18,400-12,600 K. This is shifted to
20,800-17,200 K for log $g$ = 9.0 models. In this interval of surface
gravity, the excited periods are found between $\sim$ 100 s and $\sim$
700 s, and the shorter periods are excited in the higher-gravity
models. Adding carbon in the envelope mixture tends to extinguish 
pulsational driving.}
{Given the right location in parameter space, some carbon-atmosphere white
dwarfs are predicted to show pulsational instability against gravity
modes. We are eagerly awaiting the results of observational searches for
luminosity variations in these stars.}

\authorrunning{Fontaine, Brassard, \& Dufour}
\titlerunning{Carbon-Atmosphere White Dwarfs as a New Type of Pulsator}
\maketitle

\section{Introduction}

In a recent publication, Dufour et al. (2007) reported on the
unexpected discovery of a new type of white dwarf. These are stars with
atmospheres dominated by carbon. Until then, white dwarfs cooler than
$\sim$ 80,000 K were known to come in only two flavors, those with
almost pure hydrogen atmospheres (the so-called DA stars comprising some
75 to 80\% of the known sample), and those with almost pure helium
atmospheres (the non-DA's accounting for the rest). Obviously quite rare
since none had ever been found before among the several thousands white
dwarfs known, the carbon-atmosphere stars uncovered by Dufour et
al. (2007) form a small sample of nine faint objects with $g$ $\sim$
18$-$19 culled from the Sloan Digital Sky Survey (SDSS). They are all
found in a rather narrow range of effective temperature, between 18,000
K and 23,000 K. A typical value of the abundance ratio derived from the
less noisy SDSS spectra in the sample is $X$(C)/$X$(He) $\geq$ 99. However, 
a ratio as small as $X$(C)/$X$(He) $\geq$ 3 cannot be excluded in some
of the other spectra according to the more detailed analysis reported by
Dufour et al. (2008).

Beyond their very existence, which remains to be explained (see, in this
context, the qualitative evolutionary scenario proposed by Dufour et
al. 2007), carbon-atmosphere white dwarfs have attracted our interest
from another point of view. And indeed, it has not escaped our attention 
that past models of carbon-atmosphere white dwarfs in the range of
effective temperature where the real ones are found are characterized by an
important outer superficial convection zone, very similar to that found
in the pulsating DB (V777 Her) stars (centered around $T_{\rm eff}$
$\simeq$ 25,000 K) or in the pulsating DA (ZZ Ceti) stars (found around
$T_{\rm eff}$ $\simeq$ 12,000 K). Hence, it follows that
carbon-atmosphere white dwarfs could also excite pulsation modes through
the same partial ionization/convective driving phenomenon that is at
work in these two distinct families of pulsating white dwarfs.  

In this connection, we draw attention of the reader to Figure 7 of
Fontaine \& Van Horn (1976) where some properties of the convection 
zones found in models of hydrogen-, helium-, and carbon-atmosphere white
dwarfs are compared. There are obvious similarities, and this strongly 
suggests that a stability study for carbon-atmosphere stars would be 
worth doing. The lead author would never have imagined that the
carbon envelope white dwarf models that he examined as part of his
Ph.D. thesis some 35 years ago could ever have any relevance with the
real world! At the time, this was done only out of theoretical curiosity
(see Fontaine 1974) and only the hydrogen- and helium-atmosphere models
were then thought to be ``relevant''.  And indeed, to our knowledge,    
the carbon envelope models of Fontaine \& Van Horn (1976) have remained 
the only ones of the kind to have been published.

With this background in mind, we decided to investigate the
asteroseismological potential of models of carbon-atmosphere white
dwarfs. Further impetus came from the involvement of one of us (P.D.)
with other collaborators to search observationally for luminosity
variations in these stars. We thought that it would be quite timely to
provide the theoretical background in parallel with that search.

\section{Equilibrium Models and Stability Survey}

In the sort of exploratory work presented in this paper, it is amply
sufficient to study relatively simple models of carbon-atmosphere
white dwarfs. We thus considered equilibrium structures made of a
uniform-composition envelope sitting on top of a C/O core (also with a
uniform composition specified by $X$(C) = $X$(O) = 0.5). As a working
definition of a ``carbon-dominated'' mixture, we adopted the convention
$X$(C) $\geq$ 0.5 for the mass fraction of carbon in the
envelope/atmosphere. Hence, we first constructed stellar models with an
envelope defined by $X$(C) = $X$(He) = 0.5, representing a limiting case
in the composition domain for carbon-atmosphere white dwarfs. In a
second step, and to provide a measure of the importance of varying the
carbon abundance, we also considered envelope compositions defined by
$X$(C) = 0.75 and $X$(He) = 0.25, $X$(C) = 0.9 and $X$(He) = 0.1, as
well as $X$(C) =0.99 and $X$(He) = 0.01. We explicitly ignored the
possible presence of small traces of hydrogen in our survey.

It is worth pointing out here that the white dwarf models that we used are
static, but {\sl full} stellar structures nevertheless. To compute them, 
we took advantage of the scaling law $L(r) \propto M(r)$, as appropriate 
for a cooling white dwarf (see Brassard \& Fontaine 1994 for more details 
on this). Given a composition profile, a model is further specified by
its effective temperature $T_{\rm eff}$ and surface gravity log $g$. 
Equivalently, due to the specific mass-radius relationship that
applies to white dwarfs, the total mass could be specified instead of
the surface gravity, but we prefer the first alternative because it
connects better with spectroscopy. In addition, in the range of 
effective temperature of interest here, convection is present in the
envelope of carbon-atmosphere white dwarfs (Fontaine \& Van Horn 1976),
and we described it through the so-called ML2 version of the mixing-length 
theory (see Table 1 of Fontaine, Villeneuve, \& Wilson 1981 for the 
original definition of ML2). We also briefly explored the effects of
changing the modelling of convection by computing some extra models
under the assumption of ML3 convection (a version similar to ML2 in
terms of the geometry of the eddies, but assuming a ratio of the mixing
length to the the local pressure scale height of $\alpha$ = 2 instead of
1 as in ML2).

We carried out a stability survey of these white dwarf models with the
help of our finite-element pulsation codes (Brassard et al. 1992a;
Fontaine et al. 1994). Given what is known about V777 Her and ZZ Ceti
pulsators, we specifically searched for unstable dipole ($\ell = 1$)
modes in the range of periods from 80 s to 1200 s. These would be
$g$-modes. We also concentrated on models cooler than $T_{\rm eff}$ =
30,000 K to connect with the observed carbon-atmosphere white dwarfs of
Dufour et al. (2007). We point out that we used a full nonadiabatic
approach, but within the so-called frozen convection approximation. This
approximation leads to quite reasonable estimates of the location of the
blue edge of the ZZ Ceti instability strip in the log $g$--$T_{\rm eff}$
plane (see, e.g., the discussion in Brassard \& Fontaine 1999), and the
same is true for the V777 Her instability strip (Beauchamp et al. 
1999). At the same time, it is well known that the frozen convection
approximation leads to red edges of instability strips that are cooler, 
and sometimes much cooler than they ought to. This caveat should be
kept in mind. 

A last remark about our equilibrium models concerns our choice for the
thickness of the outer carbon-dominated envelope, which we picked at a
fractional mass depth of log $q_{\rm env}$ $\equiv$ 
log $(1 - M(r)_{\rm env}/M_*)$ = $-$3.0 in all cases. At this depth, the
specified uniform envelope composition changes smoothly to the core
composition of $X$(C) = $X$(O) = 0.5 in our models. Clearly, the actual
pulsation periods must depend on this particular choice due to trapping
effects at the composition transition layer (see, e.g., Brassard et
al. 1992b). However, as discussed at some length in Quirion, Fontaine ,
\& Brassard (2004) in the context of GW Vir white dwarfs, it is not so
much the  actual periods themselves that matter in the kind of stability
study that we carried out, but the {\sl range of unstable periods}. As
explicitly demonstrated by Quirion et al. (2004), that interval of
unstable periods is very largely independent of the actual choice of 
log $q_{\rm env}$, as long as the base of the envelope is located much
deeper than the driving/damping region. The specific choice of log
$q_{\rm env}$ = $-$ 3.0 ensures just that.

\begin{figure}[!ht]
\begin{center}
\begin{tabular}{l}
\includegraphics[width=9.2cm,height=9.2cm]{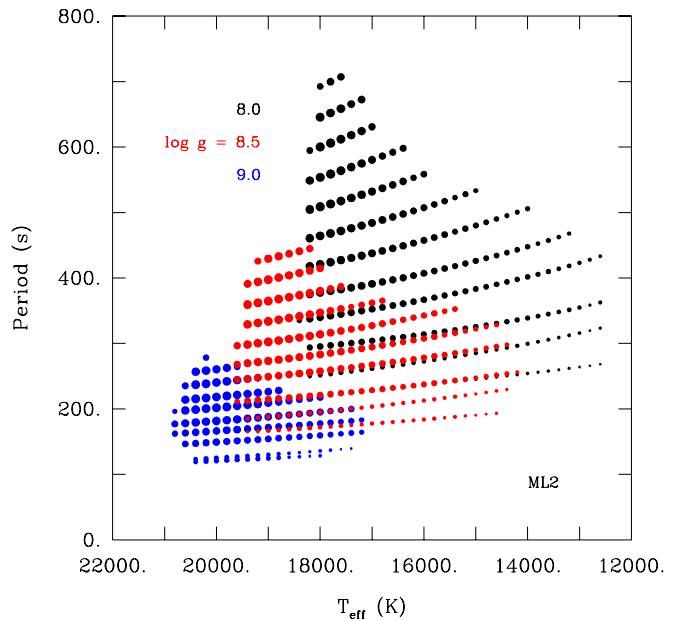}
\end{tabular}
\end{center}
\caption{Period spectra of excited dipole modes in carbon-atmosphere
  white dwarf models. Three families of models are shown differing in
  surface gravity (see the color code), but they have otherwise the same
  envelope composition ($X$(C) = $X$(He) = 0.5) and ML2 convection. Each
  dot gives the period of a mode, and its size represents a logarithmic
  measure of the modulus of the imaginary part $\sigma_I$ of the complex
  eigenfrequency. The bigger the dot, the more unstable the mode.}
\end{figure}

Figure 1 summarizes some of our results for models with an envelope
composition specified by $X$(C) = $X$(He) = 0.5 and ML2 convection. Taking 
into account the physical background briefly reported on above, we were
not really surprised, but still very much thrilled, to find out that
pulsation modes can be excited in some carbon-atmosphere white dwarf
models. This is because this finding potentially opens the door to the
application of the tools of asteroseismology for the further study of
this new class of carbon-atmosphere white dwarfs. The black dots in
Figure 1 map the expected instability strip for a reference family of
models with log $g$ = 8.0. Dipole modes in this family are excited in a
range of effective temperature from 18,400 K down to 12,600 K. For the 
reasons given above, the red edge is probably calculated cooler than it
ought to, but the blue edge should be quite secure. The excited modes
are all low-order $g$-modes, with a radial order $k$ = 2 for the lowest
series of adjacent points (separated by 200 K in $T_{\rm eff}$), to $k$ =
13 for the top three points illustrated. The total interval of excited
periods that is mapped for this family of models covers $\sim$ 200 s to
$\sim$ 700 s. We point out also that the e-folding times of the excited
modes illustrated in the figure are all much smaller than the typical
evolutionary timescale of a carbon-atmosphere white dwarf, a few times $10^8$
yrs, reaching in some cases (the largest dots in the plot) values as low
as 50 to 100 yrs. This means that such modes have plenty of time to
develop an observable anplitude.

The theoretical blue edge for the log $g$ = 8.0 family of models displayed
in Figure 1 may fall a little short if we focus, for a moment, on the
current sample of nine carbon-atmosphere white dwarfs discovered by
Dufour et al. (2007) since those cluster in a range of $T_{\rm eff}$
from 23,000 K to 18,000 K. However, a simple shift in surface gravity
easily pushes the blue edge of the instability strip well within that
range as shown by the red and blue dots in Figure 1, referring,
respectively, to a family of models with log $g$ = 8.5 and log $g$ =
9.0. Note that this displacement of the blue edge toward higher effective
temperatures is accompanied by a shift of the range of unstable periods
toward shorter values. This is not to be surprising at all because the
periods themselves (see, e.g., Brassard et al. 1992b in the case of ZZ
Ceti pulsators), i.e., those of given $k$ values, are known to decrease
when the surface gravity of a white dwarf increases. A fortiori, this is
also true for excited periods. Note further that the fact that the blue
edge temperature increases with increasing surface gravity is a
phenomenon that is also observed and well understood for the pulsating
DB (Beauchamp et al. 1999) and the pulsating DA white dwarfs (see, e.g.,
Gianninas, Bergeron, \& Fontaine 2006). 

\begin{figure}[!ht]
\begin{center}
\begin{tabular}{l}
\includegraphics[width=9.2cm,height=9.2cm]{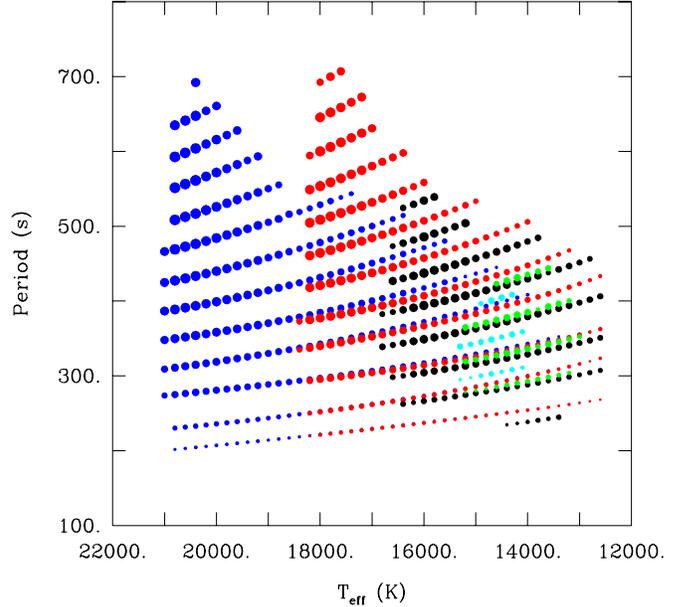}
\end{tabular}
\end{center}
\caption{Similar to Fig. 1, but the red dots refer, this time, to the
  reference family of models specified by log $g$ = 8.0, $X$(C) =
  $X$(He) = 0.5, and ML2 convection. The blue dots were obtained from
  similar models but using the more efficient ML3 version of the
  mixing-length theory. The black (green, cyan) points again refer to
  log $g$ = 8.0, ML2 models, but with an envelope composition defined by 
  $X$(C) = 0.75 and $X$(He) = 0.25 ($X$(C) = 0.9 and $X$(He) = 0.1, 
  $X$(C) = 0.99 and $X$(He) = 0.01).} 
\end{figure}

Figure 2 shows some other results of our exploratory survey, in
particular the effects of changing the chemical composition of the
envelope. The red dots correspond to our reference family of models with
log $g$ = 8.0, ML2 convection, and $X$(C) = $X$(He) = 0.5 in the
envelope. All other things being the same, the black dots show a
shrinking instability domain when the envelope composition is changed to
$X$(C) = 0.75 and $X$(He) = 0.25. There is a decrease of the blue edge
temperature from 18,400 K to 16,800 K in that case. The band of
predicted unstable dipole modes also reduces to the range $\sim$ 230 s to
$\sim$ 580 s. This trend continues to be verified in models with an 
envelope composition defined by $X$(C) = 0.9 and $X$(He) = 0.1 as
indicated by the green dots. The predicted blue edge for this family of
log $g$ = 8.0, ML2 models is now at a value $T_{\rm eff}$ = 15,300 K.
Although the location of the blue edge does not change much more, the
overall instability domain shrinks further as can be seen from the cyan 
dots associated with models with a very rich carbon composition in the 
envelope, i.e., $X$(C) = 0.99 and $X$(He) = 0.01. This is a strong
indication that the actual amount of carbon in the envelope of real
carbon-atmosphere white dwarfs will be a decisive factor in the fate of
these stars as pulsators. Finally, the blue dots in Figure 2 illustrate
the significant blueward shift (some 2,600 K) of the predicted
instability strip and some widening of the instability domain when
convection is modelled in terms of the more efficient ML3 version
compared to our standard ML2 flavor.  

\begin{figure*}[!ht]
\begin{center}
\begin{tabular}{lcr}
\includegraphics[scale=0.45,angle=0]{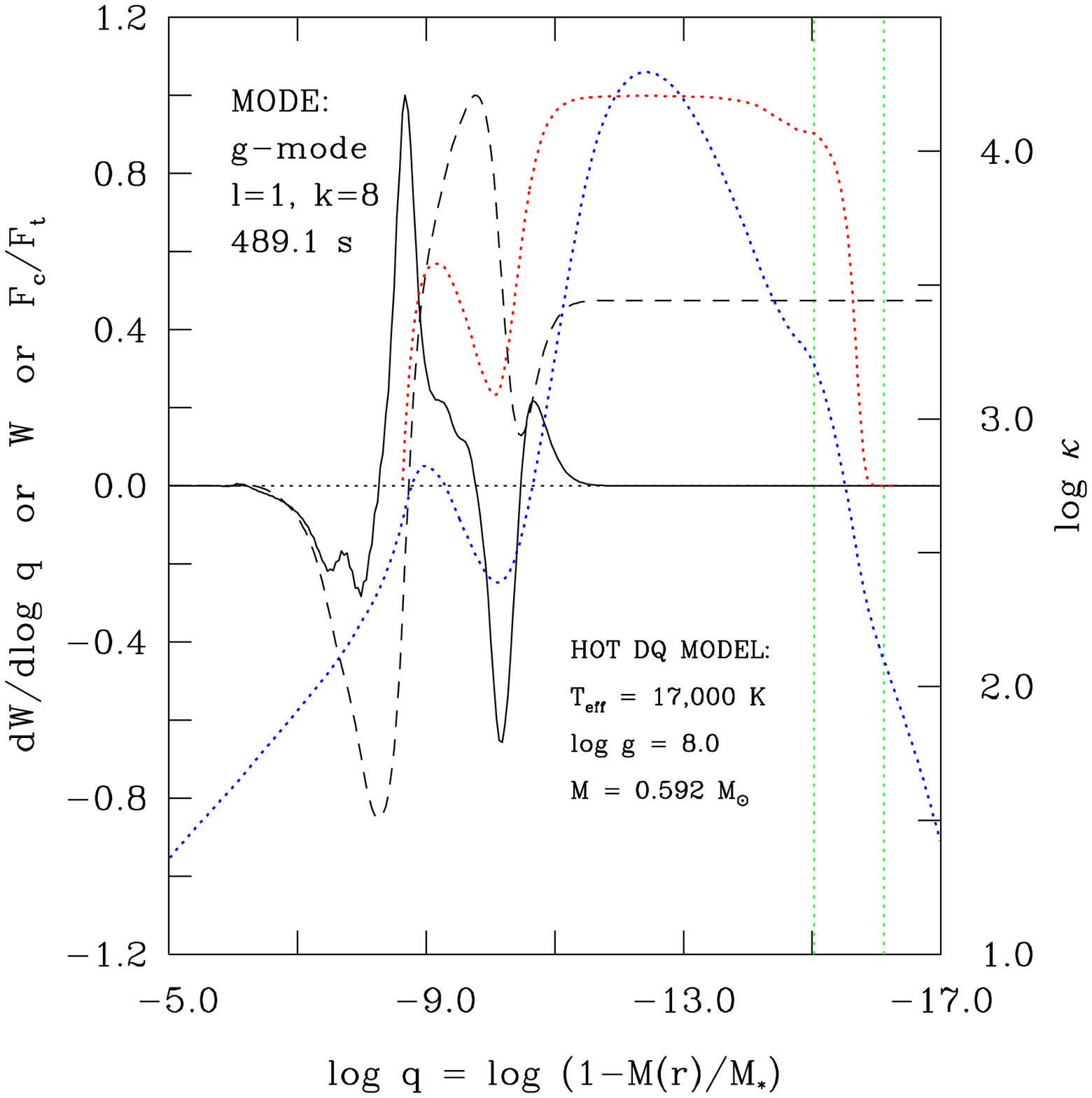}
\includegraphics[scale=0.45,angle=0]{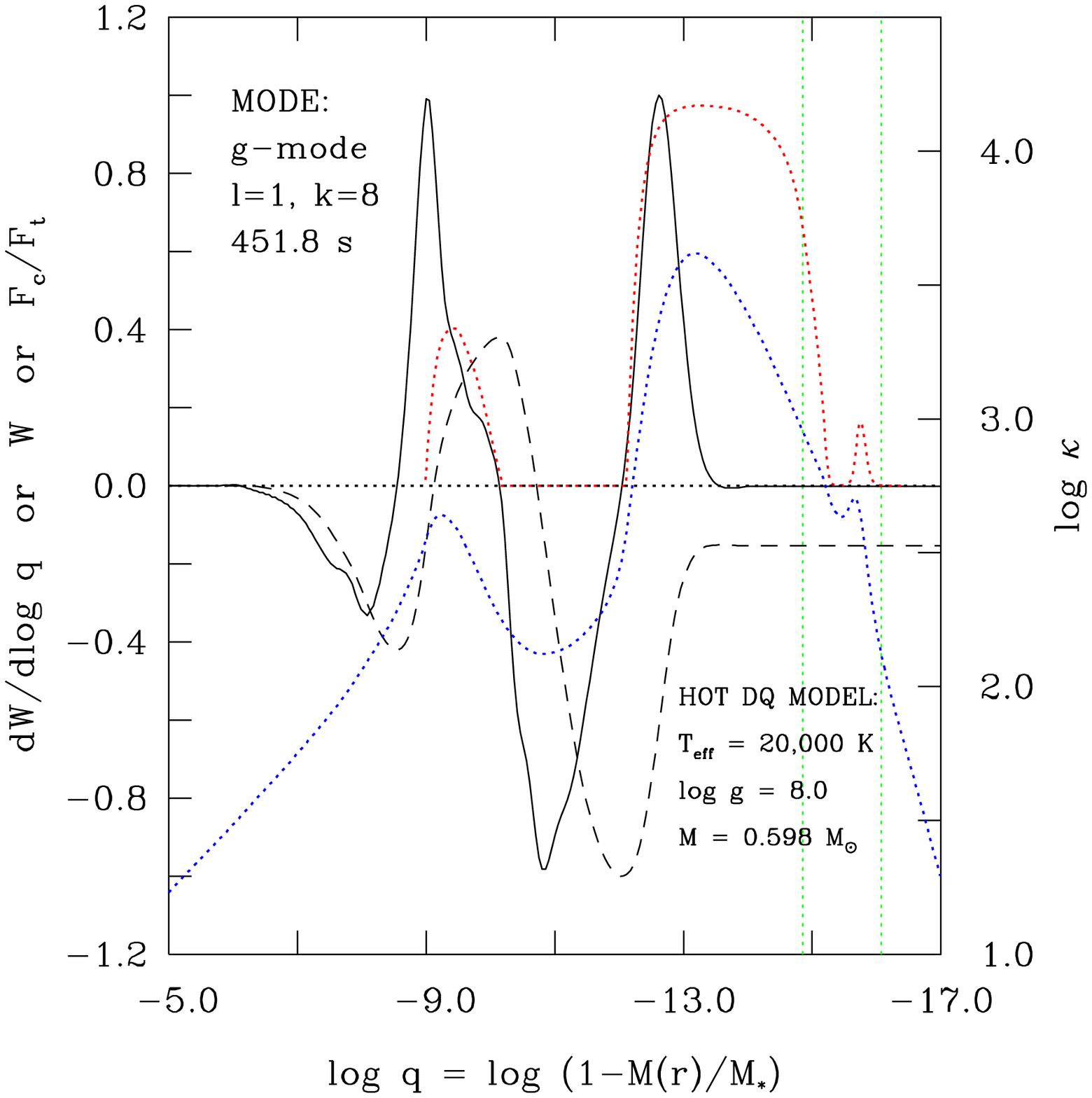}
\end{tabular}
\end{center}
\caption{\textit{Left panel}: Details of the driving/damping process for
  a typical $g$-mode excited in a 17,000 K, ML2 model of a hot DQ
  (carbon-atmosphere) white dwarf with a envelope composition $X$(C) =
  $X$(He) = 0.5 and a gravity log $g$ = 8.0. The solid curve shows
  the integrand of the work integral of the mode as a function of
  fractional mass depth. The dashed curve shows the running work
  integral, from left to right, toward the surface of the model. The red
  dotted curve shows the ratio of the convective to total flux. The blue
  dotted curve gives the run of the Rosseland opacity, to be read on
  the RHS ordinate axis. The radiative opacity values were taken from the
  OPAL data base. The maximum in the opacity profile, located at
  log $q$ $\simeq$ $-$12.41 and corresponding to a temperature $T$
  $\simeq$ $1.045 \times 10^5$ K, is caused by the partial ionization of
  He II, C III, and C IV in the envelope mixture. The secondary maximum,
  located at log $q$ $\simeq$ $-$8.98 and corresponding to a temperature $T$
  $\simeq$ $1.177 \times 10^6$ K, is caused by the partial ionization of 
  C V, and C VI. The vertical green dotted line on the left
  (right) gives the location of the base of the atmosphere at optical
  depth $\tau_R$ = 100 (of the phoptosphere at $\tau_R$ = 2/3). 
  \textit{Right panel}: Similar and for the same $g$-mode, but in
  a hotter model at 20,000 K for which all modes investigated are
  stable. It can be seen that the value of the work integral $W$ at the
  surface (dashed curve) is now negative, contrary to the case in the
  left panel, and this signifies that the mode is not excited. Note also
  the two separate convection zones in this model.} 
\end{figure*}

Given these most interesting results, we now turn briefly to the question 
of the mechanism responsible for exciting pulsation modes in these models. 
Figure 3 illustrates the details of the driving/damping region in two
models culled from the log $g$ = 8.0 family above, one ($T_{\rm eff}$ =
17,000 K) featuring excited $g$-modes, and the other ($T_{\rm eff}$ =
20,000 K) hotter than the blue edge and showing only stable modes.
Out of the many $g$-modes found excited in the 17,000 K model (left
panel), we have singled out a representative one with indices $k$ = 8
and $\ell$ = 1. It has a period of 489.1 s. We also retained the same mode,
this time stable and with a period of 451.8 s, in the 20,000 K model
(right panel). The figure illustrates a situation which bears strong
analogy with the case of the V777 Her and ZZ Ceti pulsators, but is more
complicated because of the presence of two maxima in the opacity
distribution instead of a single peak in these other pulsators. In the
present case, both opacity maxima (caused by two distinct partial
ionization zones in the envelope mixture) are ``active'' in the sense
that they both contribute to the driving/damping process. What we can
observe from the plot is that the regions on the descending side
(going in from the surface) of an opacity bump contribute locally to
driving, while the deeper adjacent zones, where the opacity plummets to
relatively low values, contribute instead to damping. In the 17,000 K
model, the two opacity bumps are relatively close to each other and are
part of a single convection zone. The damping region in between the two
bumps is then relatively narrow (see the left panel) and the overall
work integral comes out positive, meaning that the mode is globally
excited. In contrast, the two opacity maxima are relatively apart in the
20,000 K model, so much so in fact that they lead to the formation of
two distinct, separate convection zones. In that case, there is a much
larger region in between the opacity bumps where {\sl radiative} damping
becomes so large that the overall work integral comes out negative.

\section{Conclusion}

Our exploratory survey of the asteroseismological potential of
carbon-atmosphere white dwarfs has allowed us to uncover regions of
parameter space where that potential can be fulfilled. For example,
low-order $g$-modes, typical of other pulsating white dwarfs such as the
V777 Her and ZZ Ceti stars, can be excited in a wide range of effective
temperature (18,400 K to 12,600 K) for models specified by log $g$ =
8.0, ML2 convection, and an envelope composition given by $X$(C) =
$X$(He) = 0.5. Hotter models can become unstable also if their surface
gravity is increased or if the assumed convective efficiency used in the
model building phase is increased. The actual amount of carbon in the
envelope is crucial in determining the location and extent of the
instability strip. The extent of that domain tends to diminish when the
carbon abundance is increased. If real pulsating stars are found in the
current sample of carbon-atmosphere white dwarfs, we would expect them
to be massive, to show significant amounts of helium in their
atmospheres, and to be possibly undergoing efficient convective
transport.

We have provided here the basic theoretical framework within which to
interpret possible discoveries of pulsations in carbon-atmosphere white
dwarfs. We are aware of at least one observational search being
currently pursued on that, and we are eagerly awaiting the results. We
note that the sample of carbon-atmosphere stars of Dufour et al. (2007)
is small, but it is likely, now that we know how to recognize them, that
more will be uncovered from future data release from the SDSS. If real
pulsators are found among that group, they will constitute the fourth
distinct category of pulsating white dwarfs after the GW Vir, V777 Her,
and ZZ Ceti stars. More detailed pulsation studies will certainly be
warranted in that eventuality.

\end{document}